\documentclass[a4paper]{jpconf}

\usepackage{graphicx}
\usepackage{longtable}

\newcommand{\cm}{cm$^{-1}$}
\newcommand{\2}{$_{2}$}
\newcommand{\3}{$_{3}$}
\newcommand{\4}{$_{4}$}

\begin{document}
\title{Molecular line shape parameters for exoplanetary atmospheric
applications}

\author{Sergey Yurchenko, Jonathan Tennyson and Emma J. Barton}

\address{Department of Physics and Astronomy, University College London, Gower
Street, WC1E 6BT
London, UK}

\ead{s.yurchenko@ucl.ac.uk}

\begin{abstract}

We describe the recent updates to the ExoMol database
regarding the molecular spectral line shapes. ExoMol provides comprehensive
molecular line lists with a special emphasis on the applications involving
characterization of atmospheres of exoplanets and cool stars. Among important
requirements of such applications are (i) the broadening parameters for hydrogen
and helium dominating atmospheres and (ii) very broad ranges of temperature and
pressures. The current status of the available line shape data in the
literature, demands from the exoplanetary community and their specific needs are
discussed.

\end{abstract}

\section{Introduction}

Molecules in the atmospheres of exoplanets, and similar astronomical objects such as
brown dwarfs, are observed at pressures approaching atmospheric and therefore display
pronounced pressure effects in their spectra \cite{jt521,08FrMaLo}. There is, however, a severe lack
of data appropriate for reproducing these pressure effects.

The ExoMol project provide extensive line lists for a large range of molecules,
mostly related to the atmospheric retrievals for exoplanets and cool stars
\cite{jt528}. The selection of molecules and the spectroscopic coverage in
ExoMol are dictated by the requirements from these applications. A major requirement
is that the line lists must be sufficiently complete to be applicable
for high temperatures specific for atmospheres of most of these objects \cite{jt572}.
Spectroscopic data for almost
40 molecules are currently covered by the ExoMol database, including
diatomic, triatomic, tetratomic and two pentatomic species. The available line lists
were generated both by
ExoMol and other groups, see Table \ref{tab:exomoldata}.

Recently we have undertaken  a major upgrade of the ExoMol database
\cite{jt631}. The upgrade was motivated by the growing role of ExoMol as a
major provider of the hot spectroscopic line lists for atmospheric retrievals
from exoplanetary observations. The most important modifications include: (i)
introduction of application programming interface (API); (ii) line shape
parameters; (iii) cooling functions; (iv) radiative life times \cite{jt624} and Land\'{e}
g-factors \cite{jt655}. In this contribution the new line shape data provided by ExoMol are
discussed.

\section{Requirements for the line shape parameters from the exoplanetary
applications.}

It is well recognized that Voigt profiles only provide an approximate solution,
especially for high resolution atmospheric studies \cite{jt584}. However Voigt
profiles are in widespread use and are easily computed
\cite{79Hum}; they are therefore uniformly used to represent pressure broadening effects
in exoplanetary models and hence
in the ExoMol database.

Table~\ref{tab:exomoldata} illustrates the coverage of the  broadening
parameters in ExoMol, which is also a reflection of the current status in the
field. Below we list the main factors affecting the development of the line
shapes for the exoplanetary applications.

\begin{enumerate}
  \item The dominant species in gas giant planets
such as hot Jupiter atmospheres are H\2\ and He, often
at high pressure \cite{14AmBaTr.broad}. However most of the line shapes
parameters available in the literature are for the air (or N\2) as the main
broadener and motivated by the terrestrial applications. Even though the needs
for the line shapes broadened by H\2\ and He have been recently recognized by
HITRAN \cite{16WiGoKo.pb}, the corresponding line shape data is still incomplete
\cite{12Maxxxx.exo,08FrMaLo}.
  \item The large size of the line lists in ExoMol, which consist of tens of
billions of lines for larger molecules, make it practically impossible to
populate the whole database with accurate, line-by-line line shape parameters, experimental or
theoretical.  Only simple and very approximate models (e.g. based only on one
quantum number $J''$) can afford productions of the data on such a large scale.
However even simple models do not exist for most of the species in question,
especially for H\2\ and He as broadeners.
  \item  The high temperatures ($T > 1000 K$) of hot Jupiters,
and other exoplanets of current interest \cite{jt629}, require line profiles for a
large range of rotation and vibration  excitations over extended temperature ranges.
It is considered to be
important to model at least the rotational ($J$) dependence of the line widths
broadened by H\2\ \cite{14AmBaTr.broad}.
  \item There is no agreement on the value of the line wing cut-off to be used
when computing Voigt profiles. On one hand, it is imperative to use a
reasonable cut-off distance in order to speed up the line-by-line calculations
of the molecular opacities. On the other hand, the choice of the cut-off is known to effect the results
at high pressures. Common practices include a fixed cut-off (e.g. 25~\cm\ or
100~\cm\ \cite{14AmBaTr.broad}), variable cut-off depending on pressure $P$ (e.g.
min(25P; 100) \cm\ \cite{07ShBuxx.dwarfs}) or proportional to the line
(half-)width (e.g. 500 widths \cite{16HeMaxx}). The importance of folding back
the truncated wings to ensure that the strength of the profile is conserved
\cite{07ShBuxx.dwarfs} is also considered in certain circumstances.
  \item Efficient computational algorithms for the Voigt evaluations
\cite{79Hum,11Schreier} as well as the sampling method \cite{16EvSiWa.exo} are
required when billions of lines are involved.
\end{enumerate}

\begin{table}[h]
\centering
\caption{Molecular line lists included in the ExoMol
database and Voigt line shape parameters when available.}
\label{tab:exomoldata}
\begin{tabular}{lcclcc}
\hline\hline
Molecule       &  Reference                  &        Broadener         &
Molecule     &    Reference                   &        Broadener        \\
AlO            &  \cite{jt598}               &                          &
HD$^+$       &    \cite{jt506}                &                         \\
BeH            &  \cite{jt529}               &                          &
HeH$^+$      &    \cite{jt347}                &                         \\
CaH            &  \cite{jt529}               &                          &
HNO$_3$      &    \cite{jt614}                &    Air                     \\
CaH            &  \cite{11LiHaRa.cah}        &                          &    KCl
         &    \cite{jt583}                &                         \\
CaO            &  \cite{jt618}               &                          &    LiH
         &    \cite{jt506}                &                         \\
CH             &  \cite{14MaPlVa.CH}         &                          &
LiH$^+$      &    \cite{jt506}                &                         \\
CH$_4$         &  \cite{jt564}               &   Air, He, H\2, CH\4     &    MgH
         &    \cite{jt529}                &                         \\
CN             &  \cite{14BrRaWe.CN}         &                          &    MgH
         &    \cite{13GhShBe.MgH}         &                         \\
CP             &  \cite{14RaBrWe.CP}         &                          &
NaCl         &    \cite{jt583}                &                         \\
CrH            &  \cite{02BuRaBe.CrH}        &                          &    NaH
         &    \cite{jt605}                &                         \\
CS             &  \cite{jt615}               &         Air, CS          &    NH
         &    \cite{14BrRaWe.NH}          &                         \\
FeH            &  \cite{10WEReSe.FeH}        &                          &
NH$_3$       &    \cite{jt500}                &    Air, H\2, He, NH\3   \\
H$_2$CO        &  \cite{jt593}               &   Air, He, H\2, H\2CO    &
PH$_3$       &    \cite{jt593}                &    Air, He, H\2, PH\3   \\
H$_2$O         &  \cite{jt378}               &   Air, H\2, He, H\2O     &    PN
         &    \cite{jt590}                &                         \\
H$_3^+$        &  \cite{jt181}               &                          &    ScH
         &    \cite{jt599}                &                         \\
HCl            &  \cite{11LiGoBe.HCl}        &  Air, H\2, He, HCl      &    SiO
         &    \cite{jt563}                &                         \\
HCN/HNC        &  \cite{jt570}               &    Air, He, H\2, HCN     &    TiH
         &    \cite{05BuDuBa.TiH}         &                         \\
SO$_2$        &  \cite{jt635}               &    Air, He, H\2  &    SO$_3$
         &    \cite{jt641}         &                         \\	
H$_2$S        &  \cite{jt640}               &      &    H$_2$O$_2$
         &    \cite{jt638}         &                         \\	 	
VO        &  \cite{jt644}               &      &
         &            &                         \\	 	

         &                                &                         \\
\hline
\hline
\end{tabular}

\end{table}

\begin{table}[h]
\centering
\caption{Voigt parameters available in the literature and used in the ExoMol
database.}
\label{tab:broad}
\begin{tabular}{lll}
\hline\hline
Molecule  &         Broadener &   Reference    \\
\hline
H$_{2}$O  & H$_{2}$ &
\cite{14LaVoNa.h2opb,jt658,Steyert2004183,Brown1996263,05BrBeDe.h2opb,Gamache1996471,Dick2009619,jt544,Langlois1994272,93DuJoGo.h2opb,05Goxxxx.h2opb,04ZePaCo.h2opb,12DrWixx.h2opb} \\
H$_{2}$O  & He &
\cite{16PeSoSoSt.h2opb,13PeSoSoSt.h2opb,12PeSoSoSt.h2opb,08SoStxx.h2opb,09SoStxx.h2opb,14LaVoNa.h2opb,jt658,Steyert2004183,Gamache1996471,Dick2009619,95LaPnSu.h2opb,93DuJoGo.h2opb,05Goxxxx.h2opb,04ZePaCo.h2opb} \\
H$_{2}$O  & Air &
\cite{03MeJeHe.h2opb,04GaHaxx.h2opb,88GaToGo.h2opb,08PaDeCa.h2opb,05Gaxxxx.h2opb,09GaLaxx.h2opb} \\
H$_{2}$O  & H$_{2}$O &
\cite{03MeJeHe.h2opb,04GaHaxx.h2opb,94Maxxxx.h2opb,08GoKoKr.h2opb,08CaPuBu.h2opb} \\
CH$_{4}$  & H$_{2}$ &
\cite{92Pixxxx.ch4pb,93Maxxxx.ch4pb,88FoJeSt.ch4pb,93StTaCa.ch4pb} \\
CH$_{4}$  & He &
\cite{90VaChxx.ch4pb,04GaGrGr.ch4pb,01GrFiTo.ch4pb,88FoJeSt.ch4pb} \\
CH$_{4}$  & Air & \cite{09aSmBePr.ch4pb,03BrBeCh.ch4pb,08AnNiWr.ch4pb} \\
CH$_{4}$  & CH$_{4}$ & \cite{05PrBrDe.ch4pb,09bSmBePr.ch4pb,03BrBeCh.ch4pb} \\
NH$_{3}$  & H$_{2}$ & \cite{93PiMaBu.nh3pb,01HaArOr.nh3pb,07ShBuxx.dwarfs,01BaAmBu.nh3pb} \\
NH$_{3}$  & He & \cite{93PiMaBu.nh3pb,01HaArOr.nh3pb,07ShBuxx.dwarfs,01BaAmBu.nh3pb} \\
NH$_{3}$  & Air & \cite{94BrPexx.nh3pb,04NeSuVa.nh3pb,01BaAmBu.nh3pb} \\
NH$_{3}$  & NH$_{3}$ &
\cite{94BrPexx.nh3pb,Levy1993172,SergentRozey198866,Salem200423,Pickett1981197,
Levy199420} \\
PH$_{3}$  & H$_{2}$ & \cite{Bouanich2004195,Levy1993172,SergentRozey198866,Pickett1981197} \\
PH$_{3}$  & He &
\cite{Salem2005247,Levy1993172,SergentRozey198866,Pickett1981197} \\
PH$_{3}$  & Air/N$_{2}$ & \cite{Salem2005247,06BuSaKl.ph3pb,Kleiner2003293} \\
PH$_{3}$  & PH$_{3}$ & \cite{06BuSaKl.ph3pb,Kleiner2003293,89NgRoSe.ph3pb,04SaArBo.ph3pb} \\
H$_{2}$CO & H$_{2}$ & \cite{75Nexxxx.hrcopb} \\
H$_{2}$CO & He & \cite{75Nexxxx.hrcopb} \\
H$_{2}$CO & N$_{2}$ & \cite{10JaLaKw.hrcopb} \\
H$_{2}$CO & H$_{2}$CO & \cite{10JaLaKw.hrcopb} \\
SO$_{2}$ & H$_{2}$ & \cite{12CaPuxx.so2pb} \\
SO$_{2}$ & He & \cite{12CaPuxx.so2pb,14TaPiSt.so2pb} \\
SO$_{2}$ & Air & \cite{12CaPuxx.so2pb,14TaPiSt.so2pb} \\
HCN       & Air/N$_{2}$ & \cite{08YaBuGo.hcnpb,04DeBeSm.hcnpb,03RiDeSm.hcnpb} \\
HCN       & HCN & \cite{04DeBeSm.hcnpb,03DeBeSm.hcnpb} \\
HCN       & H$_{2}$ & \cite{85MeMaVr.hcnpb} \\
HCN       & He    & \cite{85MeMaVr.hcnpb} \\
HCl       & N$_{2}$  & \cite{09HuHeVa.hclpb} \\
HCl       & HCl & \cite{85PiFrxx.hclpb} \\
HCl       & H$_{2}$ & \cite{70ToHuPl.hclpb} \\
HCl       & He    & \cite{63RaEaRa.hclpb,09HuHeVa.hclpb} \\
CS        & Air & \cite{99BaWaBo.cspb} \\
CS        & CS & \cite{09MiLeBo.cspb} \\
HNO$_{3}$ & Air & \cite{89MaWexx.hno3pb} \\
\hline
\hline
\end{tabular}

\end{table}

In practice it is computationally prohibitive to perform radiative transport calculations
on hot exoplanets line-by-line. Therefore the models either use
precomputed cross sections, such as is done by $\tau$-Rex \cite{jt593,jt611}, or tables
of $k$-coefficients, as done by the NEMISIS   \cite{08IrTeKo.model}. This means that
appropriate values are precomputed on a grid of temperatures and pressures as inputs
to such codes.

\section{Conclusion}

We are in the advanced stages of developing a diet for pressure broadening
of molecules present in exoplanetary atmospheres and those of other hot astronomical
objects. A full discussion of this problem will be given elsewhere \cite{jtdiet}.

Pressure broadening data is urgently needed for atmospheric studies (retrievals)
of exoplanets and cool stars. There is huge demand on the comprehensive
solutions of the line shape problems for most of the molecules important for
exoplanetary studies. The ExoMol database is arguably the main source of opacities
for hot species important for modeling these atmospheres, where we
have created a structure for depositing and curating any molecular data
important for spectroscopic properties of hot atmospheric and other gaseous
environments. We invite the molecular data producers to contribute to this
database. The line shape data is especially important as the line parameters
are missing or incomplete for the H\2-rich atmospheres even for the most
important absorbers. Exoplanetary atmospheric retrieval is a hot topic at
the moment with a lot of interest from the society, which makes it a good place
to be for an expert in the molecular line profiles. The lack of data, a
strong demand from the field and interest from the public is a very attractive
mixture for work in this direction to be properly recognized and
rewarded.

We also invite the community to visit and test the new ExoMol database at
www.exomol.com. Any feedback will be greatly appreciated.

\section*{Acknowledgements}

This work is supported by  ERC Advanced Investigator Project 267219. We thank
the support of the COST action MOLIM (CM1405).

\section*{References}



\providecommand{\newblock}{}

\end{document}